\documentclass[superscriptaddress,nofootinbib,aps,pra,twocolumn]{revtex4-2}

\usepackage{amsmath}
\usepackage{amssymb}
\usepackage{amsfonts}
\usepackage{bm} 
\usepackage{bbm}
\usepackage{braket}
\usepackage{color}
\usepackage{comment}
\usepackage{dcolumn} 
\usepackage{dsfont}
\usepackage{enumerate}
\usepackage{epsfig}
\usepackage{esint}
\usepackage[T1]{fontenc}
\usepackage{framed}
\usepackage{textcomp, gensymb}
\usepackage{graphicx} 
\usepackage[colorlinks,linkcolor=blue,citecolor=blue,urlcolor=blue,hyperindex,driverfallback=dvipdfm]{hyperref}
\usepackage{indentfirst}
\usepackage{lmodern}
\usepackage{mathrsfs}
\usepackage{mathtools}
\usepackage{multirow}
\usepackage{psfrag}
\usepackage{pst-all}
\usepackage{soul}
\usepackage{units}
\usepackage{xcolor}
\usepackage{xspace}

\newcommand{\abs}[1] {\mathopen{}\left|#1\right|\mathclose{}}

\newcommand{\ccpar}[1] {\mathopen{}\left(#1\right)\mathclose{}}
\newcommand{\sqpar}[1] {\mathopen{}\left[#1\right]\mathclose{}}

\def\ii{{\rm i}}  \def\ee{{\rm e}}
  \def\kB{{k_{\rm B}}}
  \def\Imm{{\rm Im}}
\newcommand{\pd}[2] {\mathopen{}\frac{\partial#1}{\partial#2}\mathclose{}}

\def\rb{{\bf r}}  \def\Rb{{\bf R}}
    
\def\xx{\hat{\bf x}}  \def\yy{\hat{\bf y}}  
    
\def\kb{{\bf k}}  \def\bb{{\bf b}}
\def\Eb{{\bf E}}  \def\pb{{\bf p}}
\def\EF{{E_{\rm F}}}  
\def\eps{\epsilon}  \def\vep{\varepsilon}
\def\ww{\omega}
\def\Dm{\mathcal{D}}  \def\Hm{\mathcal{H}}
  \def\phiext{\phi^{\rm ext}}  
\def\rhoind{\rho^{\rm ind}}

\begin{document}

\title{Plasmons in phosphorene nanoribbons}

\author{Line Jelver}
\email[Line Jelver: ]{lije@mci.sdu.dk}
\affiliation{POLIMA---Center for Polariton-driven Light--Matter Interactions, University of Southern Denmark, Campusvej 55, DK-5230 Odense M, Denmark}

\author{Joel~D.~Cox}
\email[Joel~D.~Cox: ]{cox@mci.sdu.dk}
\affiliation{POLIMA---Center for Polariton-driven Light--Matter Interactions, University of Southern Denmark, Campusvej 55, DK-5230 Odense M, Denmark}
\affiliation{Danish Institute for Advanced Study, University of Southern Denmark, Campusvej 55, DK-5230 Odense M, Denmark}

\begin{abstract}
Phosphorene has emerged as an atomically-thin platform for optoelectronics and nanophotonics due to its excellent nonlinear optical properties and the possibility of actively tuning light-matter interactions through electrical doping. While phosphorene is a two-dimensional semiconductor, plasmon resonances characterized by pronounced anisotropy and strong optical confinement are anticipated to emerge in highly-doped samples. Here we show that the localized plasmons supported by phosphorene nanoribbons (PNRs) exhibit high tunability in relation to both edge termination and doping charge polarity, and can trigger an intense nonlinear optical response at moderate doping levels. Our explorations are based on a \emph{second-principles} theoretical framework, employing maximally localized Wannier functions constructed from ab-inito electronic structure calculations, which we introduce here to describe the linear and nonlinear optical response of PNRs on mesoscopic length scales. Atomistic simulations reveal the high tunability of plasmons in doped PNRs at near-infrared frequencies, which can facilitate synergy between electronic band structure and plasmonic field confinement in doped PNRs to drive efficient high-harmonic generation. Our findings establish phosphorene nanoribbons as a versatile atomically-thin material candidate for nonlinear plasmonics.
\end{abstract}

\date{\today}
\maketitle

\maketitle

\section{Introduction}

Two-dimensional (2D) materials are now ubiquitous in explorations of nanoscale light-matter interactions \cite{xia2014two,reserbat2021quantum}. Polaritons---the excitations formed when light hybridizes with collective oscillations of polarization charge---can provide extreme levels of optical confinement near and within a 2D material \cite{basov2016polaritons,gonccalves2020strong,zhang2021interface}, with 2D polariton resonances exhibiting a high sensitivity to the intrinsic electronic properties of their atomically-thin host \cite{wu2022manipulating}. These attributes have fueled intensive research efforts in the area of graphene plasmonics, which capitalizes on the intense field confinement and electrical tunability associated with plasmon resonances supported by highly-doped graphene to achieve strong light-matter interactions spanning the terahertz (THz) and infrared (IR) spectral regimes \cite{garcia2014graphene,gonccalves2016introduction}. The increased confinement offered by patterned graphene can boost plasmon resonances to near-IR frequencies, while quantum finite-size effects become prominent in the optoelectronic, thermoplasmonic, and nonlinear optical response of structures with $\lesssim10$\,nm lateral size \cite{cox2019nonlinear}.

The appealing properties of graphene plasmons have stimulated research efforts to identify other material platforms that support 2D plasmons, potentially with tunable resonances extending to near-infrared frequencies and beyond \cite{low2017polaritons,agarwal2018plasmonics,maniyara2019tunable,menabde2022image}. Few-layer black phosphorus (BP) is a low-dimensional system that is anticipated to meet these criteria \cite{low2014plasmons,debu2018tuning}: BP is an anisotropic semiconductor that exhibits high carrier mobility and a thickness-dependent band gap ranging from 0.3\,eV in the bulk to 2.0\,eV in an atomic monolayer \cite{li2014black,liu2014phosphorene,koenig2014electric}, thus presenting applications in optical switching \cite{zheng2017black}, photodetection  \cite{long2017room}, and saturable absorbers in ultrafast laser photonics \cite{li2015polarization,lu2015broadband,chen2015mechanically,guo2015black}. Phosphorene---the monolayer form of BP---exhibits a direct and appreciable band gap, in contrast to the semimetallic band structure of graphene, while also presenting an anisotropic polarizability \cite{carvalho2016phosphorene}. The plasmon resonances supported by highly-doped phosphorene are thus anticipated to exhibit strong optical confinement, electrical tunability, and directional dependence \cite{ghosh2017anisotropic,novko2021ab}.

The electronic and optical properties of phosphorene can be further controlled by patterning, as has been recently demonstrated in experimental studies of phosphorene nanoribbons (PNRs) fabricated with widths of 4-50\,nm \cite{watts2019production}. On such length scales, quantum mechanical effects should play a similarly important role in the optical response of PNRs as in narrow graphene nanoribbons \cite{thongrattanasiri2012quantum,wedel2018emergent}, while also influencing the plasmonic excitations in highly-doped PNRs that can control light on nanometer length scales. In particular, the anisotropic character of phosphorene can enhance the contrast between PNRs with different edge terminations and their associated plasmons \cite{andersen2014plasmons}. The ability to simultaneously control electronic band structure and the field enhancement provided by plasmon resonances in nanopatterned 2D materials also presents novel opportunities to control light by light in nonlinear optical applications \cite{cox2017plasmon,devega2020strong}.

Here we explore localized plasmons supported by electrically-doped phosphorene nanoribbons with lateral size of $\lesssim10$\,nm, for which finite-size and quantum mechanical effects manifest prominently in the linear and nonlinear optical response. Our theoretical description of PNRs is based on \textit{second-principles} calculations that capture salient features in the electronic band structure associated with atomic arrangements and edge structures on mesoscopic length scales. We reveal that edge states in PNRs strongly influence the threshold doping density at which plasmons emerge, beyond which plasmon resonances exhibit high tunability at near-IR frequencies and pronounced asymmetry with respect to electron and hole doping. The field enhancement supplied by plasmon resonances is found to produce comparatively large nonlinear susceptibilities---on the order of $10^{-5}$\,esu for third-order processes---while time-domain simulations further reveal efficient high-harmonic generation in doped PNRs due to the synergy between plasmonic field enhancement and electronic band structure. Our results underscore the potential that nanostructured phosphorene holds as a platform for active nanophotonics and nonlinear plasmonics.

\section{Phosphorene nanoribbons}

\begin{figure*}
    \centering
    \includegraphics[width=\textwidth]{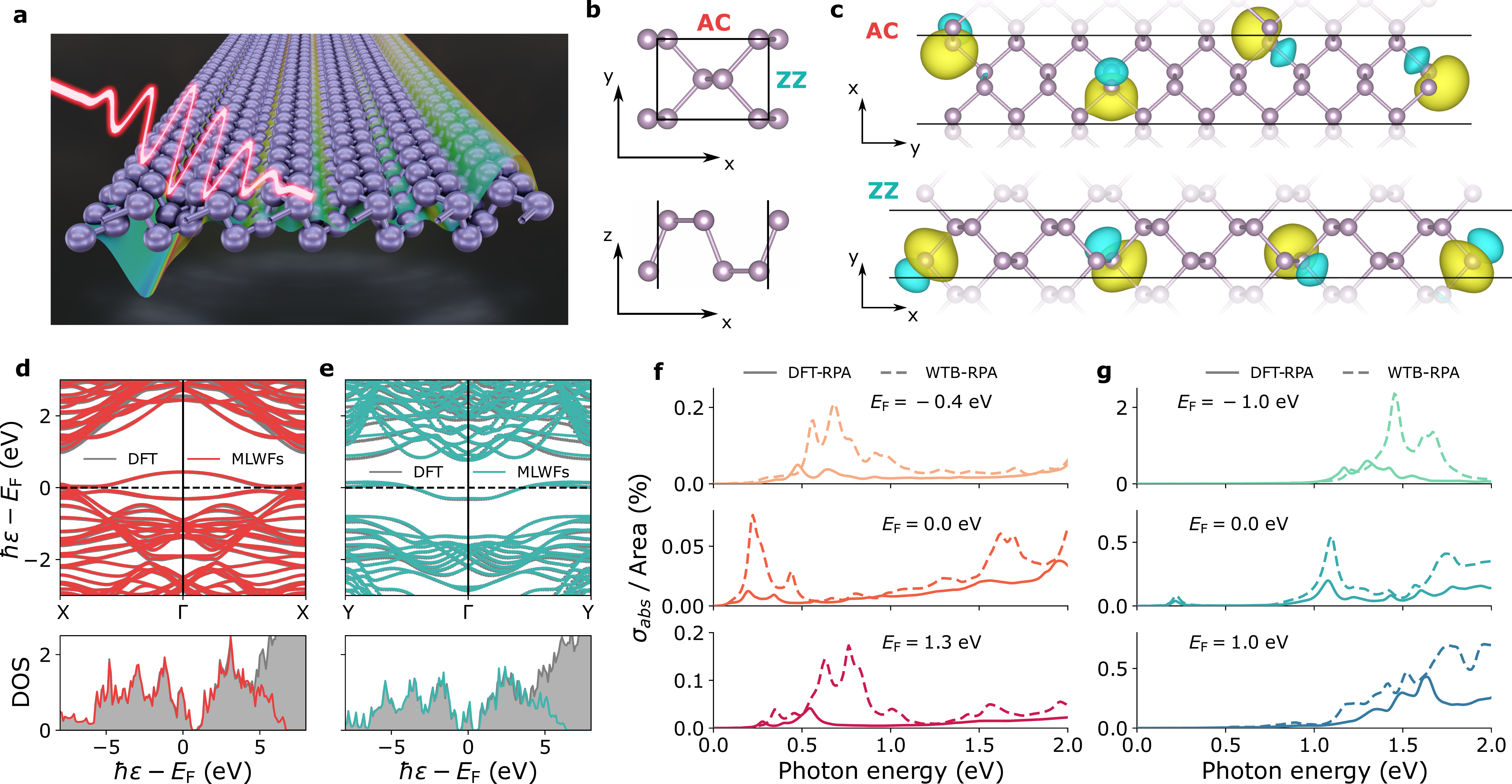}
    \caption{\textbf{Electronic and optical properties of phosphorene nanoribbons.} \textbf{a} Schematic illustration of a phosphorene nanoribbon (with armchair edge terminations and $\sim2.6$\,nm width) illuminated by light polarized in the confinement direction. \textbf{b} Top and side view of the unit cell in an extended phosphorene crystal, where AC and ZZ denote armchair and zigzag axes, respectively. \textbf{c} Top view of the unit cell an AC (top) and ZZ (bottom) edge-terminated ribbon, along with real-space maps of the four types of Wannier orbitals obtained at each atom in the two ribbons, where the yellow and turquoise colors indicate positive and negative values of the real-valued wavefunction, respectively. Panels \textbf{d} and \textbf{e} show the band structure (top panel) and density of states (DOS, bottom panel) of AC and ZZ ribbons, respectively, in each case comparing the direct DFT result to the tight-binding model constructed from MLWFs. The corresponding absorption cross-section per unit area are presented for \textbf{f} AC and \textbf{g} ZZ ribbons, calculated at the indicated Fermi energies using the Wannier tight-binding (WTB) model (solid curves) in the random-phase approximation (RPA) and compared to direct DFT calculations (dashed curves).}
    \label{fig:PNRs}
\end{figure*}

Plasmons in 2D materials exhibit an enhanced dispersion that necessitates momentum-matching schemes to in-couple light from free space \cite{gonccalves2020strong}. However, the breaking of translational symmetry through nanopatterning allows direct optical excitation of \textit{localized} plasmons: in Fig.\ \ref{fig:PNRs}a we schematically illustrate the excitation of plasmons in a highly-doped phosphorene nanoribbon (PNR) that has finite width along $\xx$ and extends infinitely along $\yy$. Phosphorene is an anisotropic 2D material that distinguishes the crystal structure along orthogonal armchair (AC) and zigzag (ZZ) axes. As illustrated schematically in Fig.\ \ref{fig:PNRs}b, these axes naturally define the unit cells of AC and ZZ PNRs shown in Fig.\ \ref{fig:PNRs}c, with distinct edge-termination and bulk atomic structure that become important in the electronic spectrum on mesoscopic length scales.

The electronic states of PNRs can be described in terms of the Kohn-Sham orbitals obtained from first-principles simulations based on density functional theory (DFT). While the electronic ground state of the PNRs are easily obtained at this level, direct atomistic simulations of the optical response become computationally prohibitive for structures with unit cells containing $\gtrsim100$s of atoms. To circumvent this limitation, we adopt a \textit{second-principles} approach based on maximally localized Wannier functions (MLWFs) \cite{marzari2012maximally}, which facilitates a tight-binding description of the Hamiltonian that can be readily adopted in optical response simulations. More specifically, we use the Quantum Espresso (QE) code \cite{QE}, invoking the Perdew-Burke-Ernzerhof (PBE) exchange correlation functional for ab-initio calculations \cite{PBE}, to create four MLWFs per phosphorous atom that capture the strong sp$_3$ type hybridization in phosophorene \cite{carvalho2016phosphorene}. These four atomic centered MLWFs are visualized in Fig.\ \ref{fig:PNRs}c for the armchair (top) and zigzag (bottom) terminated ribbons at four different atomic sites across the ribbon. Following the prescription in Methods, we obtain a Wannier tight-binding (WTB) Hamiltonian that faithfully reproduces the electronic spectra of both zigzag and armchair PNRs predicted in DFT simulations for energies within $\sim5$\,eV around the Fermi level, as shown in Fig.\ \ref{fig:PNRs}d, e.

Equipped with the WTB Hamiltonian, we compute the optical response of armchair- and zigzag-edge-terminated PNRs by adopting the single-particle density matrix formalism introduced for graphene nanostructures in Ref.\ \cite{cox2014electrically}, at the level of the random-phase approximation (RPA), with the Coulomb interaction constructed directly from Wannier orbital overlap integrals (see Methods). In Fig.\ \ref{fig:PNRs}f and g we present the absorption cross-section $\sigma^{\rm abs}=(4\pi\omega/c)\Imm\{\alpha^{(1)}\}$ obtained from the linear polarizability $\alpha^{(1)}$ for both pristine and highly-doped PNRs of $\sim2.5$\,nm width. Comparison with direct RPA simulations from full DFT calculations reveals that the WTB Hamitonian including four orbitals per atom reproduces salient features in the optical response at different doping densities, while systematically underestimating the resonant peaks. Quantitative differences in the obtained spectra could presumably arise from the more rigorous treatment of screening in DFT-RPA simulations \cite{latini2015excitons}. Indeed, as we show in the Supplementary Material, the absorption spectra calculated by neglecting Coulomb interactions in DFT-RPA and WTB-TD simulations are in excellent agreement. We further note that the single-particle formulation adopted here does not describe excitons, which should introduce narrow optical resonances near the bandgap energy and reduced by the exciton binding energy \cite{ghosh2017anisotropic}. Here, we are, however, interested in highly-doped PNRs supporting plasmon resonances formed by collective intraband excitations that are energetically below the bandgap, and should thus undergo negligible interactions with exciton resonances except in the case of very high doping where the excitons are highly damped \cite{novko2021ab}.

\section{Plasmon tunability}

\begin{figure*}[t]
    \centering
    \includegraphics[width=1\textwidth]{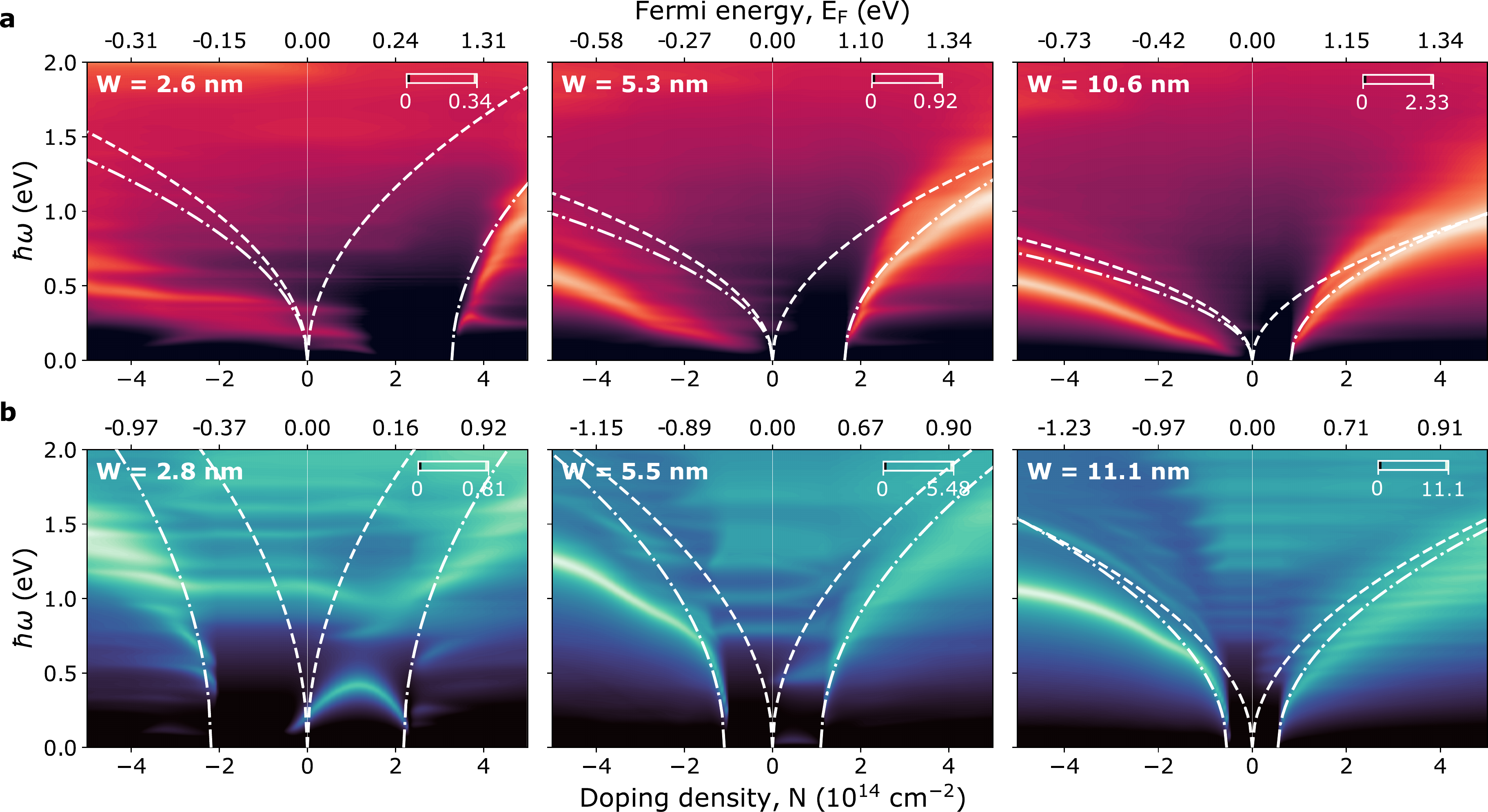}
    \caption{\textbf{Plasmon tunability in phosphorene nanoribbons.} The absorption cross-section per unit area of phosphorene nanoribbons (PNRs) is presented as a function of electron doping density $N$ and incident photon energy $\hbar\omega$ for ribbons of width $W$ and \textbf{a} armchair and \textbf{b} zigzag edge terminations. In each panel the Fermi energies reported above correspond to the specified values of $N$ on the axis ticks below, while the applied logarithmic color scale is normalized to the maximum percentage of absorption per unit area indicated by the color bar in the upper right corner. The white curves correspond to the plasmon resonances predicted by Eq. \eqref{eq:plasmons} using parameters reported for phosphorene in Ref.\ \cite{low2014plasmons} (dashed curves) or extracted from PNR electronic structure calculations while correcting for edge states (dash-dotted curves).
    }
    \label{fig:plasmons}
\end{figure*}

The plasmon resonances supported by highly-doped PNRs can be tuned by changing the ribbon width or doping charge concentration. In Fig.\ \ref{fig:plasmons} we present the absorption cross-sections predicted in WTB linear response simulations for ribbons of width $W$ as a function of the electron doping density $N$. The contour plots in Figs.\ \ref{fig:plasmons}a and \ref{fig:plasmons}b correspond to simulations of AC and ZZ PNRs, respectively, where both electron (+) and hole (-) doping up to high concentrations of $5\times10^{14}$\,cm$^{-2}$ are considered. Plasmon resonances emerge as prominent features in the absorption spectra once the doping level becomes large enough to completely fill the relatively flat bands near zero energy associated with edge states (see Fig.\ \ref{fig:PNRs}d), so that additional doping charge modifies the population in the conduction or valence band. The supported plasmons exhibit strong asymmetry in frequency and amplitude with respect to electron or hole doping, reflecting the distinct subband curvatures.

In the AC PNRs, electron doping tends to produce more intense and highly-tunable plasmon resonances owing to the steeper electron dispersion of the conduction bands. A continuum of weaker and doping-independent features corresponding to interband transitions is also observed at high energies for low doping levels, becoming quenched as the bands are populated and plasmons appear. In the ZZ PNRs, hole doping gives rise to more prominent plasmon resonances with high tunability, spanning a larger spectral range over the same doping density than their counterparts in electron-doped AC PNRs. The consistently higher energies exhibited by plasmons in hole-doped ZZ PNRs is attributed to the steeper valence band dispersion, in agreement with previous calculations on extended phosphorene \cite{agarwal2018plasmonics,prishchenko2017coulomb,novko2021ab}. Interband transitions in ZZ PNRs appear as distinct spectral features owing to the larger spacing between subbands. These results underscore the high tunability of plasmons in PNRs, as well as the possibility of strong and actively tunable plasmonic resonances in the mid-infrared spectral range, which is of great technological importance \cite{liang2021mid}. 
It is also worth noticing two opposing trends: electron doping creates the largest absorption values in a ribbon with armchair edge terminations whereas hole doping is most effective in the zigzag ribbons.

In Fig.\ \ref{fig:plasmons} we superimpose curves corresponding to the electrostatic model of plasmon resonances in nanoribbons presented in Ref.\ \cite{yu2017analytical}. For PNRs, the lowest-order dipolar plasmon resonance in the direction $\alpha=\{x,y\}$ in the case of electron (+) and hole (-) doping is given by
\begin{equation} \label{eq:plasmons}
    \ww_\alpha^\pm = \frac{1}{\sqrt{-\pi\eta_1\eps}}\sqrt{\frac{\Dm_\alpha^\pm}{W}} ,
\end{equation}
where $\eta_1$ is a negative constant which depends on the thickness $t=5.25$\,\r{A} and width $W$ of the ribbon, the former of which we take as the interlayer spacing of BP \cite{zhang2021layer}, while $\eps=1$ is the permittivity of the surrounding dielectric environment and $\Dm_\alpha^\pm$ is the Drude weight corresponding to the 2D conductivity $\sigma_\alpha^\pm(\ww) = \ii\Dm_\alpha^\pm/[\pi(\ww+\ii\tau^{-1})]$. The dashed curves in Fig.\ \ref{fig:plasmons} are plotted using the Drude weight reported in Refs.\ \cite{low2014plasmons,rodin2014strain}, which is extracted from a 4-band tight-binding model as $\Dm_\alpha^\pm=\pi e^2\abs{N}/m_{\alpha}^{\pm}$ with the effective masses $m_x^{\pm}=\hbar^2/(2\gamma_x^2/\Delta+\eta_\pm)$ and $m_y^{\pm}=\hbar^2/2\nu_\pm$ expressed in terms of parameters $\gamma_x=4a_x/\pi$\,eV\,m, $\Delta=2.0$\,eV, $\eta_{\pm}=\hbar/0.4m_0$, $\nu_+=\hbar^2/1.4m_0$, $\nu_-=\hbar^2/2.0m_0$, and $a_x=2.23$\,\r{A}.

While the $\ww_\alpha^\pm\propto\sqrt{\abs{N}}$ scaling of plasmon resonances is well-described using the parameters of Ref.\ \cite{low2014plasmons}, the band curvatures in PNRs naturally differ from those of extended phosphorene, where the absence of edge states also leads to the emergence of plasmons at arbitrarily small doping levels. The electrostatic model for the PNRs considered here can thus be improved by removing the edge state population $N\to N-4e/A$ for electron doping in AC ribbons and $N\to N\mp 2e/A$ in ZZ ribbons (the correction for holes in the former is omitted due to the overlap of edge states with the valence band), where $A$ is the area of the corresponding unit cell. The dash-dotted curves in Fig.\ \ref{fig:plasmons} show the plasmon resonances predicted by Eq.\ \eqref{eq:plasmons} using the effective masses extracted from the ab-initio electronic structure calculations performed here and implementing the edge state correction to the doping density. The electrostatic model based on bulk parameters can reasonably describe plasmon resonances in doped PNRs of $\gtrsim 5$\,nm width, while in the smallest ribbons it fails to account for important quantum finite-size effects. Furthermore, at high doping levels, kinks in the $\sqrt{N}$ scaling herald the threshold at which interband transitions redshift plasmon resonances in all PNRs. 

\section{Nonlinear response}

\begin{figure*}
    \centering
    \includegraphics[width=1\textwidth]{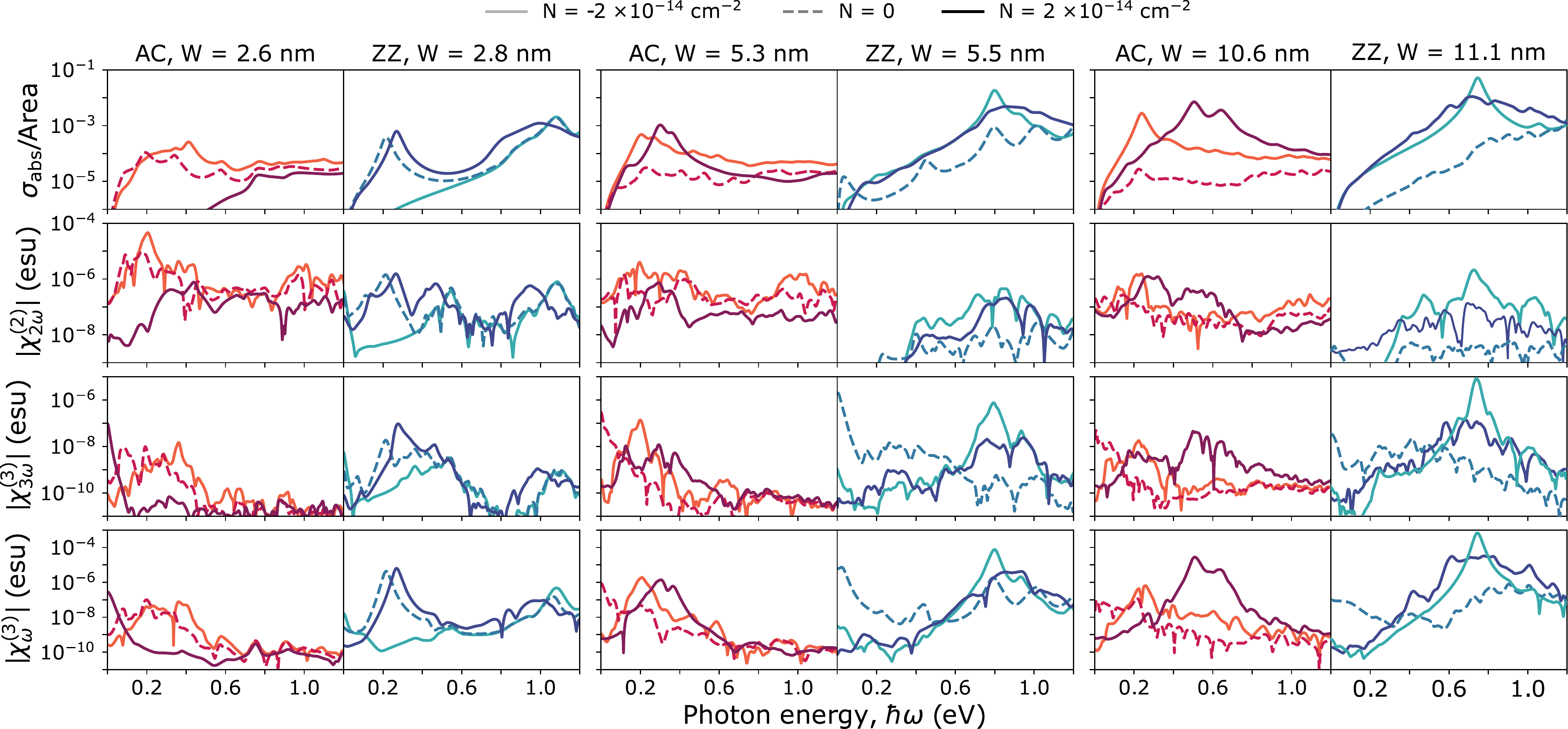}
    \caption{\textbf{Quantifying the nonlinear response of phosphorene nanoribbons.} In each column we consider armchair (AC) and zigzag (ZZ) phosphorene nanoribbons (PNRs) of width $W$ considered in Fig.\ \ref{fig:plasmons}, and present the linear absorption cross-section $\sigma^{\rm abs}$ per unit area in the upper row with the corresponding susceptibilities $\chi^{(n)}_{s\omega}$, corresponding to a nonlinear optical process at order $n$ and harmonic $s$ of the impinging frequency $\omega$, shown below. More specifically, the second and third rows correspond to second- and third-harmonic generation, respectively, while the fourth row shows the susceptibility associated with the optical Kerr effect. Results are presented for hole-doped (light curves), undoped (dashed curves), and electron-doped (dark curves) PNRs.}
    \label{fig:NL}
\end{figure*}

The tunable low-energy plasmons in highly-doped PNRs can enhance nonlinear optical processes that depend superlinearly on the impinging electric field strength. Following the procedure in Methods, we compute the nonlinear susceptibilities of PNRs in response to normally-incident monochromatic plane wave illumination $\Eb^{\rm ext}\ee^{-\ii\ww t}+{\rm c.c.}$ polarized in the ribbon confinement direction. In Fig.\ \ref{fig:NL}, we present the second- and third-order nonlinear susceptibilities of the PNRs considered in Fig.\ \ref{fig:plasmons} when the ribbons are undoped or doped to $N=\pm 2\times10^{14}$\,cm$^{-2}$.

While even-ordered nonlinear processes are forbidden in an extended phosphorene crystal due to centrosymmetry, the broken symmetry at the edges of PNRs leads to a modest second-harmonic response reaching up to $\sim10^{-5}$\,esu for the parameters considered here. As the ribbon width increases, edge effects become less significant and the nonlinear susceptibility is diminished. The second-order response is consistently higher in AC PNRs compared to their ZZ counterparts, despite the stronger plasmon resonances appearing in the linear absorption spectrum of the latter, and presumably reflecting the greater asymmetric structure of the edge terminations in AC ribbons. 

The third-order response reaches $\sim10^{-5}$\,esu in the largest PNRs, comparable to the values predicted for plasmons in highly-doped graphene nanoribbons of similar width \cite{cox2017analytical}. For the same doping density, the plasmon-assisted nonlinear optical response associated with third-harmonic generation and the Kerr nonlinearity is independent of crystal symmetry and scales with the ribbon size according to the number of free charge carriers. ZZ PNRs are found to
exhibit a consistently larger third-order response, in agreement with the trends in THG measurements performed on multilayered BP \cite{autere2017rapid,youngblood2017layer}. More specifically, the largest third-order susceptibilities are found at the plasmon resonance frequencies of the widest hole-doped PNRs, matching the relative strength of the resonances in the corresponding linear absorption spectra.

\section{High-harmonic generation}

\begin{figure}
    \centering
    \includegraphics[width=.95\columnwidth]{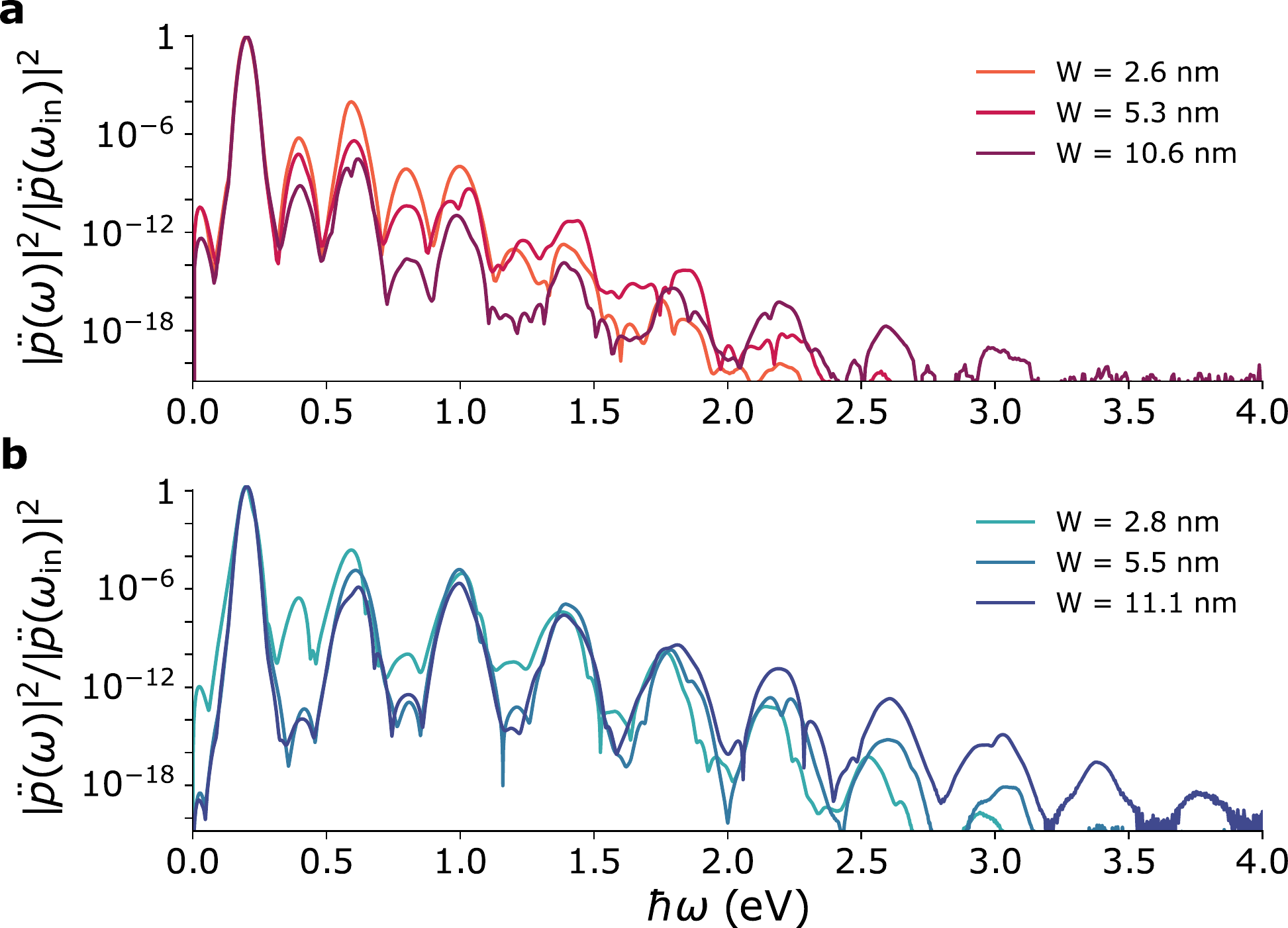}
    \caption{\textbf{High-harmonic generation in pristine phosphorene nanoribbons.} The dipole acceleration produced in undoped \textbf{a} armchair and \textbf{b} zigzag edge-terminated ribbons of width $W$ upon excitation by a Gaussian pulse with peak intensity $10^{13}$\,W/m$^2$, 100\,fs at full width half maximum duration, and 0.2\,eV central photon energy.}
    \label{fig:HHG}
\end{figure}

The excitation of condensed matter systems by intense ultrashort optical pulses can trigger high-harmonic generation (HHG), an extreme nonlinear optical phenomenon characterized by the generation of light at many integer multiples of the fundamental optical pulse carrier frequency \cite{ghimire2019high,goulielmakis2022high}. As explained in Methods, we perform time-domain simulations of the single-particle density matrix governing electron dynamics in PNRs to study their nonperturbative optical response to Gaussian pulses of 100\,fs FWHM duration. We quantify the HHG yield by the square modulus of the dipole acceleration, which is proportional to the far-field power spectrum of the emitted radiation \cite{baggesen2011dipole}. In Fig.\ \ref{fig:HHG}, we present the normalized dipole acceleration spectrum of pristine PNRs produced by a mid-infrared optical pulse with frequency $\hbar\omega_{\rm in}=0.2$\,eV (i.e., a wavelength of 6.2\,$\mu$m). Fig.\ \ref{fig:HHG}a shows the HHG yield of AC PNRs, revealing odd harmonics up to $s\sim15$ for the widest ribbon, and $\sim11$ for the narrower ribbons. In comparison, ZZ PNRs produce more prominent odd harmonics up to $s\sim19$ in the largest ribbon and $\sim15$ in the smallest ribbon. Weaker features associated with even-ordered harmonics (including a response at zero output frequency corresponding to optical rectification) also appear in the spectra, and are relatively stronger in the smallest ribbons for which edges play a more important role.

\begin{figure*}
    \centering
    \includegraphics[width=1\textwidth]{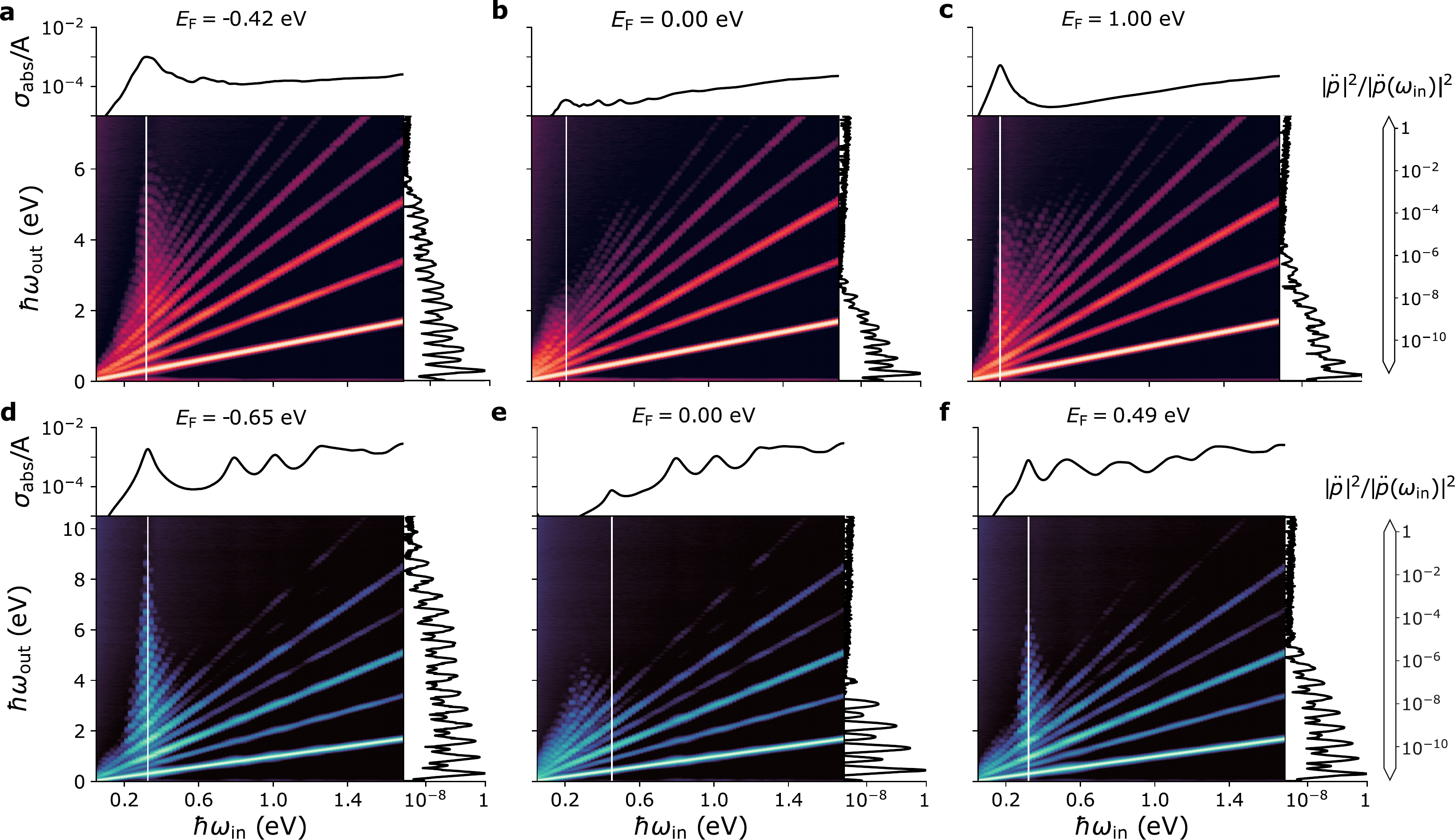}
    \caption{\textbf{Mapping high-harmonic generation in phosphorene nanoribbons.} The dipole acceleration $\ddot{p}$ produced in $W=5.3$\,nm AC (\textbf{a}-\textbf{c}) and $W=5.5$\,nm ZZ (\textbf{d}-\textbf{f}) phosphorene nanoribbons by an ultrashort optical pulse with 100\,fs FWHM duration and $10^{13}$\,W/m$^2$ peak intensity is presented as a function of input pulse energy $\hbar\omega_{\rm in}$ and the generated output energy $\hbar\omega_{\rm out}$ at various doping levels. The upper panels in each figure show the linear absorption spectra at the indicated doping level as a function of $\hbar\omega_{\rm in}$ while the right panels show the spectrum of $\abs{\ddot{p}}$ as a function of $\hbar\omega_{\rm out}$ at the input pulse energy indicated by the vertical white lines overlaying the contour.}
    \label{fig:HHG_plasmons}
\end{figure*}

Localized plasmon resonances supported by nanostructured materials present opportunities to boost light-matter interactions and enhance the HHG yield \cite{cox2017plasmon,devega2020strong}. We explore this concept in Fig.\ \ref{fig:HHG_plasmons}, where the dipole acceleration from PNRs excited by an intense ultrashort pulse is plotted as a function of the input pulse carrier energy $\hbar\omega_{\rm in}$ and the emission frequency $\hbar\omega_{\rm out}$ for doped and undoped systems. In particular, Fig.\ \ref{fig:HHG_plasmons}a-c maps the emission from AC PNRs of width $W=5.3$\,nm while in Fig.\ \ref{fig:HHG_plasmons}d-e the emission from ZZ PNRs of width $W=5.5$\,nm is presented, where in all cases the doping levels are indicated by the Fermi energy presented in the upper panel corresponding to the linear absorption cross-section of the PNR as a function of the pulse input energy. The contour plots clearly show the enhancement in HHG yield that occurs when the input pulse excites an optical resonance, a situation that is further examined in the panels on the right of each contour plot.

The results of Fig.\ \ref{fig:HHG_plasmons} indicate that hole-doped PNRs generally produce well-defined odd harmonics that persist at very high orders, e.g., up to $s\sim31$ for ZZ ribbons at the plasmon resonance. Presumably, the extended electronic band gaps exhibited by hole-doped PNRs reduces quenching of the response at the generated harmonic frequencies by interband transitions, while intraband electron motion within PNR subbands of enhanced dispersion in the valence band can give rise to efficient harmonic generation. Strong plasmon-assisted HHG from graphene nanoribbons explored using a similar level of theory in Ref.\ \cite{cox2017plasmon} was attributed to the synergy between anharmonic electron dispersion and plasmon near-field enhancement which, as we show here, also can be achieved for PNRs at comparable doping charge densities. These conclusions are also supported by a paradigmatic exploration of HHG in linear atomic chains focusing on the interplay between plasmon resonances and electronic structure, where highly-doped semiconducting chains were found to exhibit a greater HHG yield \cite{devega2020strong}. 

\section{Conclusions}

We have explored the linear and nonlinear optical response of phosphorene nanoribbons, emphasizing the enhancement in nanoscale light-matter interactions provided by localized plasmon resonances in highly-doped structures. Our investigations are based on a \emph{second-principles} theoretical description that invokes MLWFs constructed from ab-initio electronic structure calculations to treat hundreds of atoms with multiple orbitals per atom. Atomistic simulations reveal a large anistropic response in the plasmons supported by PNRs with exclusively armchair or zigzag edge terminations, which can be tuned asymmetrically by electron or hole doping. The distinction between AC and ZZ PNRs becomes even more pronounced in the plasmon-assisted nonlinear optical response, where the largest third-ordered nonlinear susceptibilities (on the order of $\gtrsim 10^{-5}$\,esu) are predicted for hole-doped ZZ PNRs. These findings are consistent with the high-harmonic generation yields estimated in non-perturbative time-domain simulations of PNRs excited by intense and ultrashort optical pulses. Our results reveal the importance of quantum finite-size effects in the linear and nonlinear optical response of plasmons supported by electrically doped PNRs, an emerging two-dimensional platform for explorations of nanoscale light-matter interactions.

\section{Methods}

The second-principles formalism introduced here to describe the optical response of phosphorene nanoribbons (PNRs) is based on DFT electronic structure calculations that are used to obtain MLWFs, which form a basis to construct tight-binding Hamiltonians and describe Coulomb interactions. These quantities form the Hamiltonian entering the equation of motion for the single-particle density matrix that governs electron dynamics in the nanoribbons, where electron-electron interactions and the coupling to an external optical field are described within the quasistatic approximation.

\subsection{Electronic states}

The electronic states of one-dimensional phosphorene nanoribbons are computed using the Quantum Espresso (QE) electronic structure code \cite{QE}, which expands the Kohn-Sham wavefunctions in a plane-wave basis set to obtain the eigenstates $\psi_{n,\kb}^{\rm KS}(\rb)$ and energies $\hbar\vep_{n,\kb}^{\rm KS}$ indexed by bands $n$ and electronic wavevectors $\kb$. We use norm-conserving pseudopotentials from the Pseudo Dojo database \cite{hamann2013optimized,van2018pseudodojo}, imposing a cutoff energy of 48\,Ry, and apply the Perdew-Burke-Ernzerhof (PBE) exchange correlation functional \cite{PBE}. We include $35$ bands per four phosphorous atoms and sample the Brillouin zone on a k-point grid of 8 points per Angstrom. The periodic 2D phosphorene crystal is allowed to relax to a force tolerance of 0.02\,eV/\r{A}$^{3}$, resulting in lattice constants $a=4.61$\,\r{A} and $b=3.34$\,\r{A} along the $\xx$- and $\yy$-directions, respectively, which define the PNRs. The electronic states of PNRs are calculated by applying periodic boundary conditions and a vacuum region of 1\,nm in both confinement directions.

Ground-state DFT using the PBE functional is known to significantly underestimate the band gap of 2D materials, as the method does not account for the highly non-local screening. For example, in extended phosphorene we obtain a bandgap of 0.89\,eV, compared to experimental measurements of $\sim$2\,eV \cite{liang2014electronic}, while a more accurate value of 2.29\,eV is obtained using a GW approach with truncated Coulomb interactions \cite{thygesen2017calculating}. However, our aim here is to reveal trends in the optical response as we change the doping, size, and edge terminations of PNRs, while plasmon resonances corresponding to collective intraband charge carrier motion are emerging at energies below the band gap, rendering the quantitative difference in band gap less important. In the case of high-harmonic generation simulations, a previous study has shown that an improvement in the bandgap value by using a meta GGA functional does not alter the HHG spectrum significantly \cite{chen2019strong}.

\subsection{Maximally-localized Wannier functions}

Maximally-localized Wannier functions (MLWFs) are constructed from the Fourier transform of the KS wavefunctions according to
\begin{equation}
    w_{l,\Rb}(\rb) = \frac{V}{(2\pi)^3}\int_{\rm BZ} d^3\kb \ee^{-\ii\kb\cdot\Rb} \sum_n \mathcal{U}_{nl,\kb}\psi_{n,\kb}^{\rm KS}(\rb) ,
\end{equation}
where $V$ is the unit cell volume, $\mathbf{R}$ is a lattice vector, and $\mathcal{U}_{jl,\kb}$ denote unitary matrices that project the Kohn-Sham states onto the Wannier basis functions at each $\kb$ \cite{marzari2012maximally}. The matrix $\mathcal{U}_{nl,\kb}$ is not uniquely defined, and is optimized for maximal spatial localization in the Wannier90 code by minimizing the gauge dependent spread
\begin{equation}
    \Omega = \sum_l\sqpar{\braket{w_{l,0}|r^2|w_{l,0}}-\rb_l^2} ,
\end{equation}
where $\rb_l\equiv\braket{w_{0,l}|\rb|w_{0,l}}$ denotes the Wannier center associated with orbital $l$ in the home unit cell ($\Rb=0$). In this work, we construct symmetry-adapted Wannier functions by restricting the Wannier orbitals to conserve the crystal symmetry during the minimization process \cite{sakuma2013symmetry}.

Once the MLWFs have been found, we construct the Wannier tight-binding (WTB) Hamiltonian at each $k$-point according to
\begin{equation}
    \Hm_{ll',k}^{\rm TB} = \sum_{m} \bra{w_{l,0}} \Hm \ket{w_{l',m}} \ee^{\ii k m b} ,
\end{equation}
such that the matrix elements $\bra{w_{l,0}} \Hm \ket{w_{l',m}}$ correspond to hopping energies between orbital $l$ in the home unit cell and orbital $l'$ in unit cell $m$ offset along the lattice vector $\bb$. Following the convention in Fig.\ \ref{fig:PNRs}, we have $\bb=b\xx$ and $\bb=b\yy$ for armchair and zigzag ribbons, respectively. The energy bands associated with the TB Hamiltonian are found by diagonalizing
\begin{equation}
    \sum_{l'} \Hm^{\rm TB}_{ll',k}a_{jl',k} = \hbar\vep_{j,k}a_{jl,k} ,
\end{equation}
where $a_{jl,k}$ are expansion coefficients for the TB eigenstates
\begin{equation}
    \ket{j,k} = \sum_{l,m}a_{jl,k}\ket{w_{l,m}}\ee^{\ii k m b} .
\end{equation}
In practice we impose an energy cutoff that neglects hoppings below 10\,meV magnitude.

The phosphorene WTB models include four MLWFs per atom and are initialized from the phosphorus s, p$_\mathrm{x}$, p$_\mathrm{y}$ and p$_\mathrm{z}$ orbitals in the case of the armchair terminated ribbons and the four sp$_3$ orbitals in the case of the zigzag terminated ribbons. The resulting Wannier orbitals reflect the strong sp$_3$ type hybridization and are similar to the orbitals shown in Fig.\ \ref{fig:PNRs}c in all the PNRs considered here. 

\subsection{Single-particle density matrix formalism}

We compute the optical response of PNRs in the quasistatic limit by adopting the formalism of Refs.\ \cite{cox2016quantum,cox2017plasmon} developed for graphene nanostructures described by a tight-binding Hamiltonian $\Hm^{\rm TB}$ interacting with a scalar potential $\phi$, for which the time-evolution of the single-particle density matrix $\rho$ is
\begin{equation} \label{eq:rho_eom}
    \pd{\rho}{t} = -\frac{\ii}{\hbar}\sqpar{\Hm^{\rm TB}-e\phi,\rho} - \frac{1}{2\tau}\ccpar{\rho-\rho^{(0)}} ,
\end{equation}
where $\rho^{(0)}$ is the equilibrium state to which the system relaxes at a phenomenological inelastic scattering rate $\tau^{-1}$. Throughout this work, we adopt a conservative scattering rate of $\hbar\tau^{-1}=50$\,meV. In the Wannier basis $\ket{l,k}\equiv\sum_j a_{jl,k}^*\ket{j,k}$, the initial state $\rho_{ll',k}^{(0)}=\sum_{j} f_{j,k} a_{jl,k}a_{jl',k}^*$ is defined by the Fermi-Dirac occupation factors $f_{j,k}=\left[\ee^{(\hbar\vep_{j,k}-\EF)/\kB T}+1\right]^{-1}$ for Fermi energy $\EF$ and electronic temperature $T=300$\,K. In practice, the Fermi energy (i.e., the chemical potential) is extracted from $N=(1/\pi)\sum_j\int_{-\pi/b}^{\pi/b}dk f_{j,k}$, where $N$ is the specified doping density.

The optical response of a PNR to a time-dependent and normally-impinging external field $\Eb(t)$ is characterized by the induced dipole moment $\pb^{\rm ind}(t) = \sum_l\rb_l\rhoind_l(t)$ per unit length $b$ along the ribbon, where $\rhoind_{l}=-(e b/\pi)\int_{-\pi/b}^{\pi/b} dk\ccpar{\rho_{ll,k}-\rho^{(0)}_{ll,k}}$ is the (spin degenerate) induced charge within the home unit cell obtained by inserting the potential
\begin{equation}
    \phi_l = \rb_l\cdot\Eb + \sum_{l'}v_{ll'}\rhoind_{l'}
\end{equation}
into Eq.\ \eqref{eq:rho_eom} and numerically evolving the density matrix elements $\rho_{ll',k}$. The first and second terms in the potential $\phi$ account for coupling to the external field and self-consistent electron-electron interactions mediated by the Coulomb operator $v_{ll'}$, respectively.

The above description of light-matter interactions assumes a negligible transfer of optical momentum along the direction of translational invariance, so that the Bloch wave vector $k$ is conserved and the induced charge $\rhoind_l$ in each unit cell is identical. The spatial dependence of the Coulomb interaction within each cell is thus also invariant, and is computed by summing over unit cells $m$ as
\begin{equation} \label{eq:coulomb}
    v_{ll'} \approx v_{ll',0} + \sum_{m\neq0}\sqpar{|\rb_l-\rb_{l'}+m\bb|^{-1}-|m\bb|^{-1}} ,
\end{equation}
where
\begin{equation}
    v_{ll',0} = \frac{1}{2} \int d^3\rb \int d^3\rb' \frac{|w_{l,0}(\rb)|^2|w_{l',0}(\rb')|^2}{|\rb-\rb'|}
\end{equation}
if $l$ and $l'$ belong to the same atom and $v_{ll',0}=|\rb_l-\rb_{l'}|^{-1}$ otherwise. In other words, we calculate the Coulomb matrix elements between orbitals of the same atom in the home unit cell from the MLWFs following the procedure in the Supplementary Material, while imposing a $1/r$ dependence for inter-atom interactions. Note that in Eq.\ \eqref{eq:coulomb} we have safely subtracted the diverging part of the Coulomb interaction that ultimately must vanish to maintain charge neutrality \cite{thongrattanasiri2012quantum}.

\subsection{Perturbative expansion}

To quantify the nonlinear response associated with specific processes, Eq.\ \eqref{eq:rho_eom} can be solved perturbatively for monochromatic illumination $\Eb(t) = \Eb^{\rm ext}\ee^{-\ii\ww t}+{\rm c.c.}$ by expanding the density matrix $\rho(t) = \sum_{n,s}\rho^{n s}\ee^{-\ii s\ww t}$ in order $n$ and harmonic $s$ of the external field amplitude. The corresponding induced charge $\rho_l^{ns}\equiv-(e b/\pi)\int_{-\pi/b}^{\pi/b}dk\,\rho_{ll,k}^{ns}$ is obtained by solving the self-consistent equation
\begin{equation} \label{eq:rho_ns}
    \rho^{ns}_{l} = \sum_{l'}\chi_{ll'}^{(0)}(s\ww)\phi_{l'}^{ns} + \frac{b}{2\pi}\int_{-\pi/b}^{\pi/b}dk\sum_{jj'}a_{jl,k}a^*_{j'l,k}\eta_{jj',k}^{ns} ,
\end{equation}
where the noninteracting susceptibility
\begin{align} \label{eq:chi0}
    \chi_{ll'}^{(0)}(s\ww) = &\frac{2e^2}{\hbar}\frac{b}{2\pi}\int_{-\pi/b}^{\pi/b}dk  \\
    &\times\sum_{jj'}\ccpar{f_{j',k}-f_{j,k}}\frac{a_{jl,k}a_{j'l,k}^*a_{jl',k}^*a_{j'l',k}}{s\ww+\ii/2\tau-\ccpar{\vep_{j,k}-\vep_{j',k}}}  \nonumber
\end{align}
mediates the self-consistent potential
\begin{equation}
    \phi_l^{ns}=\rb_l\cdot\Eb^{\rm ext}\delta_{n,1}\ccpar{\delta_{s,-1}+\delta_{s,1}} + \sum_{l'}v_{ll'}\rho^{ns}_{l'}
\end{equation}
containing the external field only at linear order, while the quantity
\begin{equation}
    \eta_{jj',k}^{ns} = \frac{2e^2}{\hbar}\sum_{n'=1}^{n-1}\sum_{s'=-n'}^{n'}\sum_{ll'}\frac{\ccpar{\phi_l^{n's'}-\phi_{l'}^{n's'}}a_{jl,k}^*a_{j'l',k}}{s\ww+\ii/2\tau-(\vep_{j,k}-\vep_{j',k})}\rho_{ll',k}^{n-n',s-s'}
\end{equation}
contributes to the nonlinear ($n>1$) response through the mixing of lower-order contributions.

\subsection{First principles optical response calculations}

The first-principles calculations of the absorption spectra shown in Fig.\ \ref{fig:PNRs}f,g are performed using the PBE functional in the GPAW electronic structure code \cite{mortensen2005real,enkovaara2010electronic}, from which Kohn-Sham eigenstates are obtained in a plane-wave basis by implementing the projector-augmented wave method with a cutoff energy of 300\,eV and the same number of bands and k-point density as in the QE calculations. In the Supplementary Material we provide additional computational parameters, along with a comparison between the DFT electronic bands obtained in QE and GPAW calculations that shows their agreement. Using the GPAW code, the noninteracting susceptibility $\chi^{(0)}$ in the form of Eq.\ \eqref{eq:chi0} is constructed directly from the Kohn-Sham eigenstates rather than the Wannier functions, and leads to a similar linear response as that predicted using the WTB Hamiltonian in the absence of electron-electron interactions, i.e., by replacing $\phi\to\phiext$ in Eq.\ \eqref{eq:rho_ns} (see SM). The interacting suceptibility is computed from a Dyson-like equation utilizing a truncated Coulomb interaction in the GPAW code, as described in Refs.\ \cite{yan2011linear,latini2015excitons}, from which the linear polarizability is obtained.

\section{Acknowledgements}

The authors thank C.~Wolff for fruitful discussions and a shared enthusiasm for the application of Wannier functions. J.~D.~C. is a Sapere Aude research leader supported by VILLUM FONDEN (grant no. 16498) and Independent Research Fund Denmark (grant no. 0165-00051B). The Center for Polariton-driven Light--Matter Interactions (POLIMA) is funded by the Danish National Research Foundation (Project No.~DNRF165). Computation in this project was performed on the DeiC Large Memory HPC system managed by the eScience Center at the University of Southern Denmark.


%

\end{document}